\documentclass[10pt,twocolumn,preprint]{emulateapj}
\usepackage[english]{babel}
\usepackage{amsmath,xcolor,aas_macros,ulem}

\newcommand{\bs}{\boldsymbol}

\newcommand{\Msun}{M_\odot}

\newcommand{\tcv}{\textcolor{violet}}

\newcommand{\tcx}{\textcolor{teal}}

\shortauthors{Yu et al.} \shorttitle{CUBE}
\begin{document}
\title{CUBE: An Information-optimized parallel Cosmological $N$-body Algorithm}
\author{
Hao-Ran Yu$^{1,2,3\star}$,
Ue-Li Pen$^{2,1,4,5,6}$,
Xin Wang$^{2}$
}

\affil{
$^1$Tsung-Dao Lee Institute, Shanghai Jiao Tong University, Shanghai, 200240, China\\
$^2$Canadian Institute for Theoretical Astrophysics, University of Toronto, Toronto, ON M5H 3H8 Canada\\
$^3$Department of Astronomy, Shanghai Jiao Tong University, Shanghai, 200240, China\\
$^4$Dunlap Institute for Astronomy and Astrophysics, University of Toronto, Toronto, ON M5S 3H4, Canada\\
$^5$Canadian Institute for Advanced Research, CIFAR Program in Gravitation and Cosmology, Toronto, ON M5G 1Z8, Canada\\
$^6$Perimeter Institute for Theoretical Physics, Waterloo, ON N2L 2Y5, Canada
}

\email{$\star$ haoran@cita.utoronto.ca}

\begin{abstract}
Cosmological large scale structure $N$-body simulations are computation-light, memory-heavy problems in supercomputing. The considerable amount of memory is usually dominated by an inefficient way of storing more than sufficient phase space information of particles. We present a new parallel, information-optimized, particle-mesh-based $N$-body code CUBE, in which information-efficiency and memory-efficiency are increased by nearly an order of magnitude. This is accomplished by storing particle's relative phase space coordinates instead of global values, and in the format of fixed point as light as 1 byte. The remaining information is given by complementary density and velocity fields (negligible in memory space) and proper ordering of particles (no extra memory). Our numerical experiments show that this information-optimized $N$-body algorithm provides accurate results within the error of the particle-mesh algorithm. This significant lowering of the memory-to-computation ratio breaks the bottleneck of scaling up and speeding up large cosmological $N$-body simulations on multi-core and heterogeneous computing systems.

\end{abstract}


\maketitle

\section{Introduction}\label{s.intro}
The $N$-body simulation, a dynamical simulation of a group of particles, is a powerful tool in physics and astronomy \citep{1988csup.book.....H}. It is widely used in cosmology to model the large scale structure (LSS) of the universe \citep{1985ApJ...292..371D}. Current percent and sub-percent level LSS measurements of cosmological parameters, via the matter power spectrum \citep{2005MNRAS.360L..82R,2011ApJ...726....7T}, baryonic acoustic oscillations (BAO) \citep{2005ApJ...633..560E,2009ApJ...700..479T}, weak gravitational lensing \citep{2003ApJ...592..699V,2009A&A...499...31H,2009ApJ...701..945S} etc., require understandings of the nonlinear dynamics of the cosmic structure, and rely on high-resolution and high dynamic range $N$-body simulations.

When $N$ is large, the brute force pairwise particle-particle (PP) force brings unaffordable $o(N^2)$ computations, so many algorithms were designed to alleviate it. Various fast-multipole methods \citep{1985JCoPh..60..187R,2014ComAC...1....1D,2016ascl.soft09016P} improve the complexity to $o(N\log N)$ even $o(N)$, among which the most popular one is ``tree'', like {\tt GADGET} \citep{2001NewA....6...79S,2005MNRAS.364.1105S} and its simulation ``Millennium'' \citep{2005Natur.435..629S,2012MNRAS.426.2046A}, {\tt TPM} \citep{1995ApJS...98..355X} and {\tt GOTPM} \citep{2004NewA....9..111D}. Other methods include adaptive grid algorithms like {\tt HYDRA} \citep{1995ApJ...452..797C} and {\tt RAMSES} \citep{2010ascl.soft11007T}, as well as mesh-refined codes \citep{1991ApJ...368L..23C} and moving adaptive particle-mesh (PM) codes \citep{1995ApJS..100..269P}. The standard PM algorithm \citep{1988csup.book.....H} is most memory and computational efficient if we focus on large cosmological scales. The load-balancing problem is minimized because the matter distribution is rather homogeneous, and the speed benefits from the fast Fourier transform (FFT) libraries, such as {\tt FFTW3} \citep{Frigo05thedesign}. {\tt PMFAST} \citep{2005NewA...10..393M} introduces a 2-level PM algorithm, aiming to push PM codes toward speed, memory compactness, and scalability. After subsequent developments on {\tt PMFAST}, {\tt CUBEP3M} \citep{2013MNRAS.436..540H} uses cubic spatial decomposition, and adds PP force and many other features.

In addition to the new methodology, the fast development of parallel supercomputers enables us to simulate a system of more than a trillion ($10^{12}$) $N$-body particles. To date, the largest $N$-body simulation in application is the ``TianNu'' \citep{2017NatAs...1E.143Y,2017RAA....17...85E} run on the TianHe-2 supercomputer. With the code {\tt CUBEP3M} adding neutrino modules, it uses $3\times 10^{12}$ particles to simulate the cold dark matter (CDM) and cosmic neutrino evolution through the cosmic age.

Relative to the optimized computation optimizations, $N$-body simulations use a considerable amount of memory to store the information of particles. Their phase space coordinates $(x,y,z,v_x,v_y,v_z)$ are stored as at least six single-precision floating point numbers (a total of 24 bytes). On the other hand, modern supercomputer systems use multi-cores, many integrated cores (MIC) and even densely parallelized GPUs, bringing orders of magnitude higher computing power, whereas these architectures usually have limited memory allocation. Thus, these computation-light but memory-heavy applications, compared to matrix multiplication and decomposition calculations, are currently less suitable for full usage of the computing power of modern supercomputers. For example, although native and offload modes of {\tt CUBEP3M} are able to run on the Intel Xeon-PHI MIC architectures, with the requirement of enough memory, TianNu simulations were done on TianHe-2 using only its CPUs -- 73\% of the total memory but only 13\% of the total computing power. We investigate how the greatest amount of information on particles can be deduced while still preserving the accuracy of $N$-body simulations, with the aim of optimizing the total memory usage for a given $N$.

In the following sections we present a new, information-optimized parallel cosmological $N$-body simulation code {\tt CUBE} \citep{2018ascl.soft05018Y}, using as little as 6 bytes per particle (bpp). It gives accurate results in cosmological LSS simulations -- the error induced by information optimization is below the error from the PM algorithm. In section \ref{s.method} we show how the memory can be saved by using an ``integer-based storage'' and how the PM $N$-body algorithm is adapted with this storage format. In section \ref{s.results} we quantify the accuracy of this algorithm using groups of simulations from {\tt CUBEP3M} and {\tt CUBE}. Discussions and conclusions are provided in section \ref{s.discussion}.

\section{Method}\label{s.method}
The most memory-consuming part of an $N$-body simulation is usually the phase space coordinates of $N$-body particles -- 24 bpp -- which contains 6 single-precision floating numbers that must be used to store each particle's 3-dimensional position and velocity vectors. {\tt CUBEP3M}, an example of a memory-efficient parallel $N$-body code, can use as little as 40 bpp when sacrificing computing speed \citep{2013MNRAS.436..540H}. This includes the phase coordinates (24 bpp) for particles in the physical domain and buffered region, a linked list (4 bpp), and a global coarse mesh and local fine mesh. Sometimes 4-byte real numbers are not necessarily adequate in representing the {\it global} coordinates in simulations. If the box size is many orders of magnitude larger than the interactive distance between particles, especially in the field of resimulation of dense subregions, double-precision (8 byte) coordinates are needed to avoid round-off errors. Another solution is to record {\it relative} coordinates for both position and velocity. {\tt CUBE} replaces the coordinates and linked list 24+4=28 bpp memory usage with an integer-based storage, thus reducing the basic memory usage from 28 bpp down to 6 bpp, as described in \ref{ss.position} and \ref{ss.velocity}. The algorithm is described in \ref{ss.overview}.

\subsection{Particle position storage}\label{ss.position}
We construct a uniform mesh throughout the space and each particle belongs to its parent cell of the mesh. Instead of storing the global coordinates of each particle, we store its offset relative to its parent cell that contains the particle. This is similar to storing the quantities of nodes/clumps (structures of a tree in a tree code) relative to their parents \citep{1985SJSSC...6...85A}. We divide the cell, in each dimension $d$, evenly into $2^8=256$ bins, and use a 1 byte (8 bits) integer $\chi_d \in \{-128,-127,...,127\}$ to indicate which bin it locates in this dimension. The global locations of particles are given by a cell-ordered format in memory space, and a complementary number count of particle numbers in this mesh (density field) will give complete information on the particle distribution in the mesh. Then, the global coordinate in the $d$th dimension $x_d$ is given by $x_d=(n_c-1)+(\chi_d+128+1/2)/256$, where $n_c=1,2,...,N_c$ is the index of the coarse grid. The mesh is chosen to be coarse enough such that the density field takes negligible memory. This coarse density field can be further compressed into a 1-byte integer format, such that a 1-byte integer shows the particle number in this coarse cell in a range from 0 to 255. In the densest cells (this rarely happens) where there are $\ge 255$ particles, we can just write 255, and write the actual number as a 4-byte integer in another file.

In a simulation with volume $L^3$ and $N_c^3$ coarse cells, particle positions are stored with a resolution of $L/(256N_c)$. The force calculation (e.g. softening length) should be configured to be much finer than this resolution, as discussed in later sections. On the other hand, particle position can also be stored as 2-byte (16 bits) integers to increase the resolution. In this case, each coarse cell is divided into $2^{16}=65536$ bins and the position resolution is $L/(65536N_c)$, which is precise compared to using 4-byte global coordinates (see later results). We denote this case as ``x2'' and denote the case in which we use 1-byte integers for positions as ``x1''.

We collectively write the general position conversion formulae as
\begin{equation}\label{eq.chi}
	\chi_d=\left[2^{8n_\chi}(x_d-\left[x_d\right])\right]-2^{8n_\chi-1},
\end{equation}
\begin{equation}\label{eq.x}
	x_d=(n_c-1)+2^{-8n_\chi}\left(\chi_d+2^{8n_\chi-1}+1/2\right),
\end{equation}
where $[\ ]$ is the operator to take the integer part. $n_\chi\in\{1,2\}$ is the number of bytes used for each integer, and $x_d$ and $\chi_d$ are floating and integer versions of the coordinate. The velocity counterparts of them are $n_\nu=1,2$, $v_d$ and $\nu_d$.
The position resolution for an $n_\chi$-byte integer, ``x$n_\chi$'', is $2^{-8n_\chi}L/N_c$.

As a $n_\chi=1$, 1D ($d=1$), 4-coarse-cell ($N_c=4$) example, if $$\chi_1=(-128,127,0,60),$$ and particle number density $${\rho_c^{\rm 1D}}=(1,0,2,1),$$ then in units of coarse cells, the accurate positions of these four particles are $$x_1=(0.001953125, 2.998046875, 2.501953125, 3.736328125).$$

\begin{figure}
\centering
  \includegraphics[width=1\linewidth]{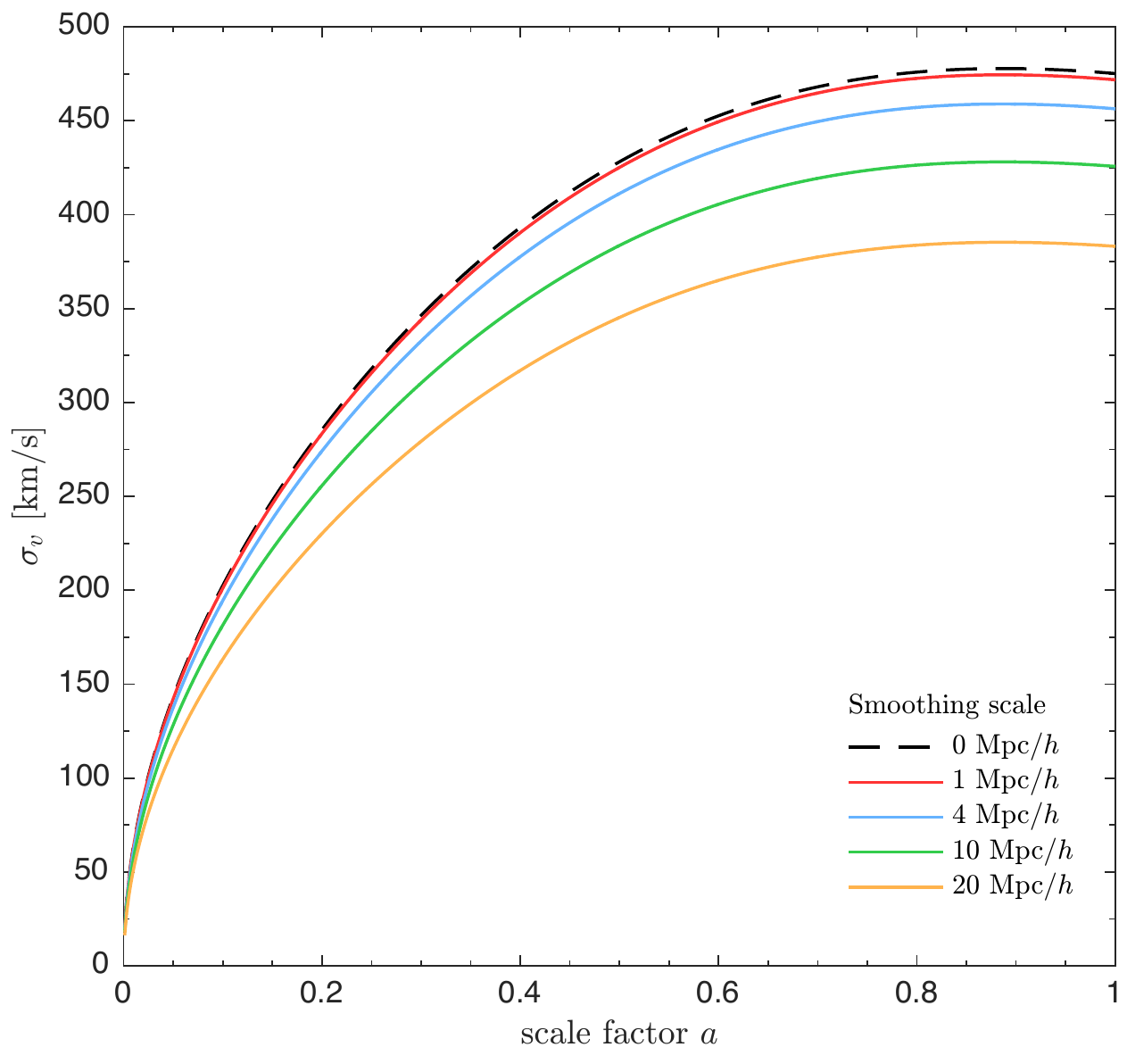}
 \caption{Variance of the velocity distribution given by Equation \ref{eq.vdisp}. The $x$-axis is the scale factor characterizing the cosmic evolution and the $y$-axis shows the $\sigma_v$ in units of km/s. The 5 curves, from top to bottom, show the $\sigma_v(a)$ with increasing smoothing scale $r$ from 0 to 20 Mpc$/h$.}
\label{fig.vdisp}
\end{figure}

\subsection{Particle velocity storage}\label{ss.velocity}
Similarly, the actual velocity in the $d$th dimension $v_d$ is decomposed into an averaged velocity field on the same coarse grid $v_c$ and a residual $\Delta v$ relative to this field:
\begin{equation}\label{eq.deltav}
	v_d=v_c+\Delta v.
\end{equation}
$v_c$ is always recorded and kept updated, and should not occupy considerable memory. We then divide velocity space $\Delta v$ into uneven bins, and use a $n_\nu$-byte integer to indicate which $\Delta v$ bin the particle is located.

The reason why we use uneven bins is that slower particles are more abundant compared to faster ones, and one should better resolve slower particles by tracing at least linear evolution. On the other hand, there could be extreme scattering particles (in case of PP force), and we can safely ignore or less resolve those nonphysical particles. One of the solutions is that, if we know the probability distribution function (PDF) $f(\Delta v)$ we divide its cumulative distribution function (CDF) $F(\Delta v)\in(0,1)$ into $2^{8n_\nu}$ bins to determine the boundary of $\Delta v$ bins, and particles should evenly distribute in the corresponding uneven $\Delta v$ bins. Practically we find that either $f(v_d)$ or $f(\Delta v)$ is close to Gaussian, so we can use Gaussian CDF, or any convenient analytic functions that are close to Gaussian, to convert velocity between real numbers and integers.

The essential parameter of the velocity distribution is its variance. On a nonlinear scale, the velocity distribution function is non-Gaussian. However, to the first-order approximation, we simply assume it as Gaussian and characterized it by the variance
\begin{equation}\label{eq.vdisp}
	\sigma^2_v(a,r) = (a H f D)^2 \int_0^\infty d^3k\frac{P(k)}{k^2}W^2(k,r),
\end{equation}
where $a(z)$ is the scale factor, $H(z)$ is the Hubble parameter, $D$ is the linear growth factor, $f=d \ln D/d\ln a$, and $P(k)$ is the linear power spectrum of density contrast at redshift zero. $W(k,R)$ is the Fourier transform of the real space top-hat window function with a smoothing scale $r$. In Figure \ref{fig.vdisp} we plot $\sigma_v(a,r)$ as a function of $a$ for a few smoothing scale $r$. $\Delta v$ in equation (\ref{eq.deltav}) is the velocity dispersion relative to the coarse grid, so we approximate its variance as
\begin{equation}\label{eq.vdelta}
	\sigma^2_{\Delta}(a)=\sigma^2_v(a,r_c)-\sigma^2_v(a,r_p),
\end{equation}
where $r_c$ is the scale of the coarse grid, and $r_p$ is the scale of average particle separation. In each dimension of the 3D velocity field, we use $\sigma^2_{\Delta}(a)/3$ according to the equipartition theorem. On different scales, we measure the statistics of $v_d$, $v_c$ and $\Delta v$ and find good agreement with the above model.

\begin{figure}
\centering
  \includegraphics[width=0.9\linewidth]{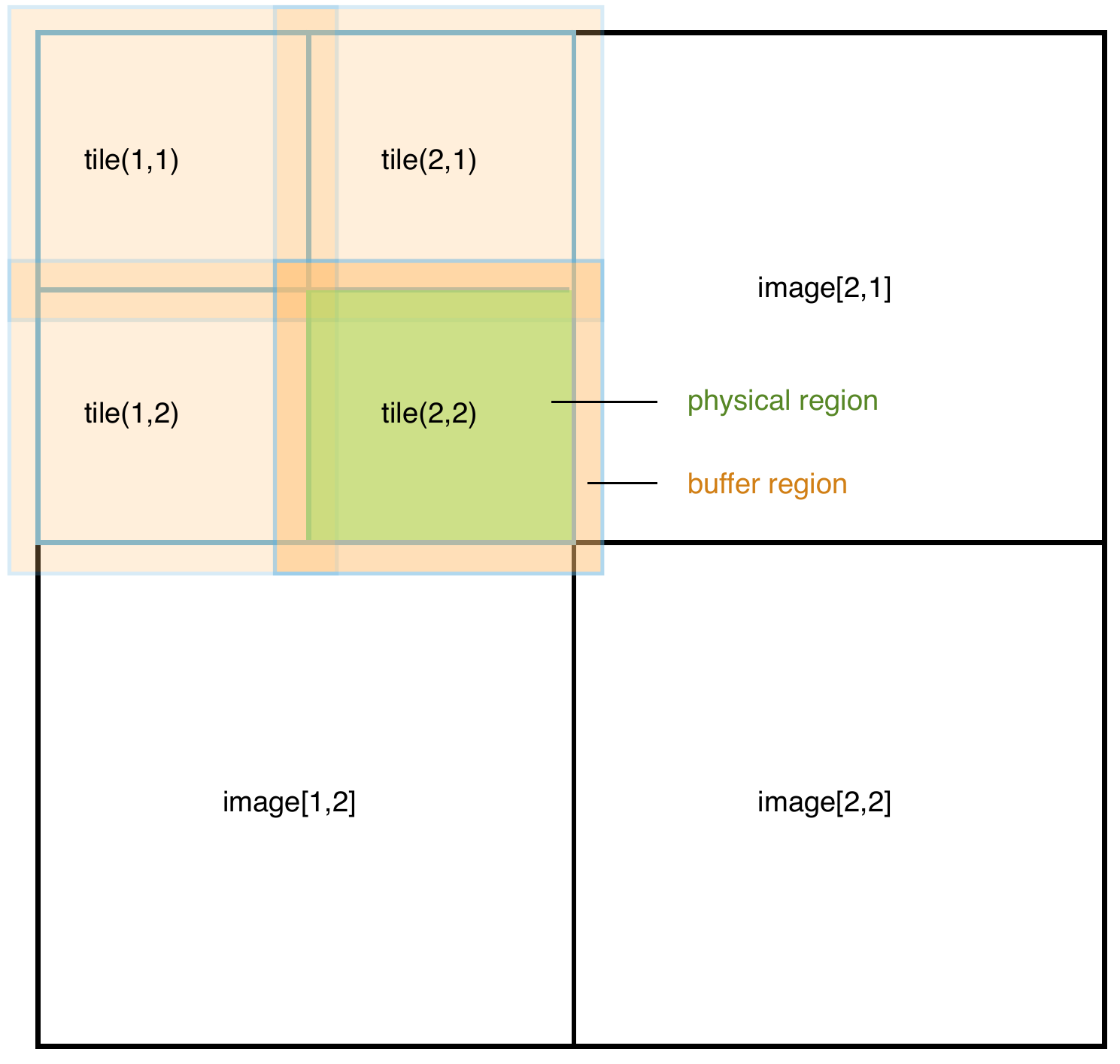}
 \caption{Spatial decomposition in {\tt CUBE} in a 2D analogy. In this example, there are two images per dimension ($M_g=2$), and two tiles per image per dimension ($M_t=2$). The orange boxes show the overlapped extended($V_e$)$\,=\,$physical($V_p$)$\,\cup\,$buffer($V_b$) tile regions. One physical region is indicated in green.}
\label{fig.tile}
\end{figure}

The simulation results are very insensitive if we manually tune the variance of the model $\sigma_\Delta$ within an order of magnitude. However, in the $n_\nu=1$ case, the method of using uneven bins gets much better results than simply using equal bins between minimum and maximum values $[\min(\Delta v),\max(\Delta v)]$. So, one can safely use a standard $\Lambda$CDM (cold dark matter with a cosmological constant as dark energy) for slightly different cosmological models, in equation (\ref{eq.vdisp}). In {\tt CUBE}, the velocity conversion takes the formula
\begin{equation}\label{eq.nu}
	\nu_d=\left\lfloor(2^{8n_\nu}-1)\pi^{-1}\tan^{-1}\left((v_d-v_c)\sqrt{\pi/2\sigma_\Delta^2}\right)\right\rfloor,
\end{equation}
\begin{equation}\label{eq.v}
	v_d=v_c+\tan\left(\frac{\pi\nu_d}{2^{8n_\nu}-1}\right)\sqrt{2\sigma_\Delta^2/\pi},
\end{equation}
where $\lfloor\ \rfloor$ is the operator to take the nearest integer. Tangent functions are convenient and computing-efficient. Compared to the error functions used in the Gaussian case, they take the same variance at $\nu_d=0$ but resolve high velocities relatively better.  All possible choices of conversion formulae and $\sigma_\Delta$ are unbiased in the conversion Equations (\ref{eq.nu},\ref{eq.v}), however, a proper choices optimize the velocity space sampling and can result in more precise results.

Initially, particles are generated by an initial condition generator, at a higher redshift. The coarse grid velocity field $v_c$ is also generated at this step by averaging all particles in the coarse cell. A global $\sigma_\Delta$ is calculated by equation (\ref{eq.vdelta}), where linear approximation holds. Then, velocities are stored by equation (\ref{eq.nu}). During the simulation, $v_c$ is updated every time-step, and a nonlinear $\sigma_\Delta$ is measured directly from the simulation, and can be simply used in the next time step, after scaled by the ratio of growth factors between two adjacent time-steps. For more details see section \ref{sss.algorithm}.

\subsection{Code overview}\label{ss.overview}
{\tt CUBE} uses a 2-level PM force calculation. In order to apply the integer-based format to the $N$-body simulation, substantial structural changes need to be done. {\tt CUBE} is written in Coarray Fortran, where Coarray features replace MPI (Message Passing Interface) communications between computation nodes/images\footnote{Images are the concept of computing nodes or MPI tasks in Coarray Fortran. We use this terminology in this paper.}. The algorithm is described in this language.

\subsubsection{Spatial decomposition}\label{sss.spatial_decomposition}
{\tt CUBE} decomposes the global simulation volume into $M_g^3$ cubic sub-volumes with $N_c$ coarse grids or $N_f=RN_c$ fine grids per side. The fine mesh is usually $R=4$ times finer than the coarse mesh. Each of these sub-volumes is assigned to a coarray {\it image}. Inside of an image, the sub-volume is further decomposed into $M_t^3$ cubic {\it tiles} (defined as $V_p$, or a {\it physical} region) with $N_c/(M_g M_t)$ coarse grids per side. Each $V_p$ is surrounded by a {\it buffer} region $V_b$ which is $N_b$ coarse cells thick. We define the extended tile region as $V_e\equiv V_p\cup V_b$. $V_e$ is designed for two purposes:

(1) If the short fine mesh force ${\bs F}_f$ has a cut-off, $N_b$ ${\bs F}_f(r>N_b)=0$, and is computed on $V_e$, then ${\bs F}_f$ in $V_p$ is guaranteed to be correct. 

(2) $V_e$ is able to collect all particles that are able to travel to $V_p$.

Figure \ref{fig.tile} shows the spatial decomposition in a 2-dimensional analogy, with $M_g=2$ and $M_t=2$.

According to this spatial decomposition, and as discussed in the last two subsections, we declare $\{\rho_c,v_c,\chi_d,\nu_d\}$ by using Fortran language:
\begin{eqnarray*}
	&&\tcv{{\tt integer(1)}}\ \ \, \rho_c(N_e,N_e,N_e,M_t,M_t,M_t)[M_g,M_g,*]\\
	&&\tcv{{\tt real(4)}}\ \ \ \ \ \ \ \, v_c(N_e,N_e,N_e,M_t,M_t,M_t)[M_g,M_g,*]\\
	&&\tcv{{\tt integer(}n_\chi{\tt )}}\, \chi_d(3,P_{\rm max})[M_g,M_g,*]\\
	&&\tcv{{\tt integer(}n_\nu{\tt )}}\ \nu_d(3,P_{\rm max})[M_g,M_g,*]
\end{eqnarray*}
where $N_e=N_t+2N_b$ covers the buffer region on both sides, $M_t$ is the tile dimensions, and $M_g$ is the image {\it co-dimensions}.\footnote{Coarray Fortran concept. Co-dimensions can enable communications between images.} We denote the actual number of particles in a given image as $P_{\rm local}$, and $P_{\rm max}>P_{\rm local}$ is a value (discussed in Section \ref{ss.memory}) large enough to store particles in $V_e$. $\chi_d$ and $\nu_d$
must be sorted according to the same memory layout as $\rho_c$, such that $n_c$ and $x_d$ can be obtained from equation (\ref{eq.x}).
$\{\rho_c,v_c,\chi_d,\nu_d\}$ provides a complete information on the positions and velocities of particles, and we call it a checkpoint.

An additional particle-ID (PID) array, $I_P$, can also be declared to differentiate particle types (e.g. CDM and neutrino particles) or to differentiate every particle, by using an $n_I$-byte integer per particle. If PID is turned on, the array $I_P$ is also included in the checkpoints, and the ordering of $I_P$ is the same as that for $\chi_d$ and $\nu_d$.

\begin{figure}[t]
{\tt \tcv{program} Initial\_Condition\_Generator\_for\_CUBE\\
\indent \ \ calculate $\Phi$ in Fourier space\\
\indent \ \ \tcv{do} (each tile)\\
\indent \ \ \ \ \tcv{do} (each particle at ${\bs q}\in V_e$)\\
\indent \ \ \ \ \ \ ${\bs \Psi}({\bs q})=\nabla\Phi({\bs q})$\\
\indent \ \ \ \ \ \ $\{{\bs q},{\bs\Psi({\bs q})}\}\rightarrow\{{\bs x},{\bs v}\}$\\
\indent \ \ \ \ \ \ $\{{\bs x},{\bs v}\}\rightarrow \{\rho_c,v_c\}$\\
\indent \ \ \ \     \tcv{enddo}\\
\indent \ \ \ \ \tcv{do} (each particle at ${\bs q}\in V_e$)\\
\indent \ \ \ \ \ \ ${\bs \Psi}({\bs q})=\nabla\Phi({\bs q})$\\
\indent \ \ \ \ \ \ $\{{\bs q},{\bs\Psi({\bs q})}\}\rightarrow\{{\bs x},{\bs v}\}$\\
\indent \ \ \ \ \ \ calculate particle's index $i$ according to $\rho_c$\\
\indent \ \ \ \ \ \ ${\bs x}\rightarrow\chi_d(:,i);\,\{{\bs v},v_c\}\rightarrow\nu_d(:,i)$\\
\indent \ \ \ \ \ \ \tcv{if} (PID\_flag) create $I_P(i)$\\
\indent \ \ \ \ \tcv{enddo}\\
\indent \ \ \ \ \tcv{do} (each coarse grid $\in V_e$)\\
\indent \ \ \ \ \ \ delete particles $\in V_b$\\
\indent \ \ \ \ \tcv{enddo}\\
\indent \ \ \ \ write $\{\rho_c,v_c,\chi_d,\nu_d\}$ to disk\\
\indent \ \ \ \ sum up $P_{\rm local}$\\
\indent \ \ \tcv{enddo}\\
\indent \ \ \tcv{sync all}\\
\indent \ \ sum up $P_{\rm global}$\\
\tcv{end}\\}
\caption{Pseudocode for the initial condition generator.}
\label{fig.ic}
\end{figure}

\subsubsection{Initial conditions}\label{sss.ic}
The cosmological initial condition generator is compiled and run separately from the main $N$-body code. Here, we briefly describe it for completeness.

The first step is the calculation of the displacement potential $\Phi$. At an initial redshift $z_i$, we generate a linear density fluctuation $\delta_L({\bs q})$ on Lagrangian grid ${\bs q}$ by multiplying a Gaussian random field with the transfer function $T(k)$ (given by the assumed cosmological model) in Fourier space. Then, we solve for the potential ${\Phi({\bs q}, z_i)}$ of a curlless displacement field ${\bs\Psi({\bs q})}$ by Poisson equation $-\nabla^2\Phi=\delta_L$ in Fourier space.

The second step to generate particles and displace them by Zel'dovich approximation (ZA) \citep{1970A&A.....5...84Z}, where the displacement field is obtained by differentiating $\Phi$, is ${\bs \Psi}({\bs q})=\nabla\Phi({\bs q})$ in real space. This step is done on $V_e$ of each tile. 

We iterate twice over particles' ${\bs q}$ in $V_e$. The first iteration calculates particles' ${\bs x}$ and ${\bs v}$ by ZA and obtains $\rho_c$ and $v_c$ on the coarse grid. The second iteration's ${\bs x}$ and ${\bs v}$ are calculated again and are converted to $\chi_d$ and $\nu_d$ by Equations (\ref{eq.chi},\ref{eq.nu}) and placed in a certain order according to $\rho_c$. Lastly, we delete particles in $V_b$, and re-sort the ones in $V_p$ and write $\{\rho_c,v_c,\chi_d,\nu_d,I_P({\rm optional})\}$ of this tile to disk. The above is similar to {\tt \tcx{update\_xp}} of Section \ref{sss.algorithm}. If PIDs are needed, they are also generated here.
After working on all tiles, we sum up $P_{\rm local}$ and $P_{\rm global}$. We summarize the above steps into a pseudocode in Figure \ref{fig.ic}. During this step, the only major memory usage is $\Phi$ on the fine mesh. If the number of particles per fine grid $P_f=1$, the memory consumption of this in-place FFT is 4 bpp.

\begin{figure}[]
{\tt \tcv{program} CUBE\\
\indent \ \ \tcv{call} \tcx{initialize}\\
\indent \ \ \tcv{call} \tcx{read\_particles}\\
\indent \ \ \tcv{call} \tcx{buffer\_density}\\
\indent \ \ \tcv{call} \tcx{buffer\_xp}\\
\indent \ \ \tcv{call} \tcx{buffer\_vp}\\
\indent \ \ \tcv{do}\\
\indent \ \ \ \ \tcv{call} \tcx{timestep}\\
\indent \ \ \ \ \tcv{call} \tcx{update\_xp}\\
\indent \ \ \ \ \tcv{call} \tcx{buffer\_density}\\
\indent \ \ \ \ \tcv{call} \tcx{buffer\_xp}\\
\indent \ \ \ \ \tcv{call} \tcx{update\_vp}\\
\indent \ \ \ \ \tcv{call} \tcx{buffer\_vp}\\
\indent \ \ \ \ \tcv{if}(checkpoint\_step) \tcv{then}\\
\indent \ \ \ \ \ \ \tcv{call} \tcx{update\_xp}\\
\indent \ \ \ \ \ \ \tcv{call} \tcx{checkpoint}\\
\indent \ \ \ \ \ \ \tcv{if} (final\_step) \tcv{exit}\\
\indent \ \ \ \ \ \ \tcv{call} \tcx{buffer\_density}\\
\indent \ \ \ \ \ \ \tcv{call} \tcx{buffer\_xp}\\
\indent \ \ \ \ \ \ \tcv{call} \tcx{buffer\_vp}\\
\indent \ \ \ \ \tcv{endif}\\
\indent \ \ \tcv{enddo}\\
\indent \ \ \tcv{call} \tcx{finalize}\\
\tcv{end}\\}
\caption{Overall structure of {\tt CUBE}. Sections of the code are grouped into Fortran subroutines, which are described in paragraphs of Section \ref{sss.algorithm}.}
\label{fig.code}
\end{figure}

\subsubsection{Algorithm}\label{sss.algorithm}
Figure \ref{fig.code} shows the overall structure of the main code.

{\tt \tcx{initialize}} creates fine mesh and coarse mesh FFT plans, and reads in configuration files telling the program at which redshifts we need to do checkpoints, halofinds, or stop the simulation. Force kernels $K_c$, $K_f$ are also computed or loaded.

{\tt \tcx{read\_particles}}, from the disk, reads in a checkpoint $\{\rho_c,v_c,\chi_d,\nu_d,I_P({\rm optional})\}$ for each image. Because they exist only in the $V_p$ of every tile, they are {\it disjoint}; and they provide {\it complete} information on the whole simulation volume -- we call it ``{\it disjoint state}''. In this state, $\rho_c$, $v_c$'s values in buffer regions, and $\chi_d(:,P_{\rm local}+1:)$ and $\nu_d(:,P_{\rm local}+1:)$ are 0's. Because $I_P$ is generated and manipulated together with $\nu_d$, so we do not explicitly mention $I_P$ in the followings.

{\tt \tcx{buffer\_density}}, {\tt \tcx{buffer\_x}} and {\tt \tcx{buffer\_v}} convert the ``disjoint'' state to the ``{\it buffered state}''. In {\tt \tcx{buffer\_density}}, $V_b$ regions of $\rho_c$ are synchronized between tiles and images. By {\tt \tcx{buffer\_x}}, $\chi_d$ is updated to contain common, buffered particles, and they are sorted according to the buffered $\rho_c$. {\tt \tcx{buffer\_v}} deals with $\nu_d$ in a similar manner.

{\tt \tcx{timestep}} is
a second order Runge-Kutta method is used in the time integration, i.e., for $n$ time-steps, we update positions ($D$=drift) and velocities ($K$=kick) at interlaced half time-steps by operator splitting; the operation $({\rm DKKD})^n$ is second-order accurate. The actual simulation applies varied time-steps by {\tt \tcx{timestep}}, where a time increment ${\rm d}t$ is constrained by particles' maximum velocities, accelerations, cosmic expansion, and any other desired conditions.

\begin{figure}[t]
{\tt \tcv{subroutine} \tcx{update\_xp}\\
\indent \ \ \tcv{do} (each physical tile)\\
\indent \ \ \ \ \tcv{do} (each particle)\\
\indent \ \ \ \ \ \ $\{\rho_c,\chi_d\}\rightarrow {\bs x};\,\{v_c,\nu_d\}\rightarrow {\bs v}$\\
\indent \ \ \ \ \ \ ${\bs x}={\bs x}+{\bs v}\,{\rm d}t$\\
\indent \ \ \ \ \ \ update $\rho_c^*,\,v_c^*$ according to ${\bs x}$\\
\indent \ \ \ \ \tcv{enddo}\\
\indent \ \ \ \ \tcv{do} (each particle)\\
\indent \ \ \ \ \ \ $\{\rho_c,\chi_d\}\rightarrow {\bs x};\,\{v_c,\nu_d\}\rightarrow {\bs v}$\\
\indent \ \ \ \ \ \ ${\bs x}={\bs x}+{\bs v}\,{\rm d}t$\\
\indent \ \ \ \ \ \ calculate particle's index $i$ according to $\rho_c^*$\\
\indent \ \ \ \ \ \ ${\bs x}\rightarrow\chi_d^*(:,i);\,\{{\bs v},v_c\}\rightarrow\nu_d^*(:,i)$\\
\indent \ \ \ \ \tcv{enddo}\\
\indent \ \ \ \ \tcv{do} (each coarse grid)\\
\indent \ \ \ \ \ \ discard buffer information\\
\indent \ \ \ \ \tcv{enddo}\\
\indent \ \ \ \ replace $\{\rho_c,v_c,\chi_d,\nu_d\}$ with $\{\rho_c^*,v_c^*,\chi_d^*,\nu_d^*\}$\\
\indent \ \ \tcv{enddo}\\
\indent \ \ \tcv{sync all}\\
\indent \ \ update velocity dispersion $\sigma^2_{\Delta}$\\
\indent \ \ sum up $P_{\rm local},\,P_{\rm global}$\\
\tcv{end}\\}
\caption{Pseudocode for subroutine {\tt \tcx{update\_xp}}.}
\label{fig.update_xp}
\end{figure}

{\tt \tcx{update\_xp}}
is used, according to ${\rm d}t$, to update particle positions (drift $D$) in a ``{\it gather}'' algorithm tile by tile. For each particle, $\chi_d$ and $\nu_d$ are converted to $x_d$ and $v_d$ by Equations(\ref{eq.x},\ref{eq.v}).

In order to keep particles ordered, for each tile, we first perform $x_d=x_d+v_d\,{\rm d}t$ on all particles to obtain an updated density and velocity field on the tile,  $\rho_c^*$ and $v_c^*$. Then, this calculation is done on the same tile again to generate a new, local particle list $\chi_d^*$ and $\nu_d^*$ by Equations(\ref{eq.chi},\ref{eq.nu}). Here, the ordering of  $\chi_d^*$ and $\nu_d^*$ relies on  $\rho_c^*$. Then, the third iteration is done on this tile to delete buffer regions of  $\{\rho_c^*,v_c^*,\chi_d^*,\nu_d^*\}$.  Then the disjoint state of $\{\rho_c^*,v_c^*,\chi_d^*,\nu_d^*\}$ replaces the old $\{\rho_c,v_c,\chi_d,\nu_d\}$. Finally, $P_{\rm local}$ and $P_{\rm global}=\sum P_{\rm local}$ are updated. These steps are summarized in Figure \ref{fig.update_xp}.

In {\tt \tcx{update\_vp}}
the PM or PP particle-mesh (P$^3$M) algorithm is applied in this subroutine to update particles' velocities (kick $K$).
We first call {\tt \tcx{buffer\_density}} and {\tt \tcx{buffer\_xp}} to place the particle positions in the buffered state. Then, according to the particle distributions $\rho_f$ on $V_e$, we calculate the fine mesh force ${\bs F}_f$ and update the particle velocities in the $V_p$. An optional PP force ${\bs F}_{\rm pp}$ (subroutine {\tt \tcx{PP\_force}}) can be called to increase the force resolution.

\begin{figure}[t]
{\tt \tcv{subroutine} \tcx{update\_vp}\\
\indent \ \ \tcv{do} (each extended tile)\\
\indent \ \ \ \ \tcv{call} \tcx{PP\_force}\\
\indent \ \ \ \ calculate $\rho_f$ and solve ${\bs F}_f$ in Fourier space\\
\indent \ \ \ \ \tcv{do} (each particle in physical region)\\
\indent \ \ \ \ \ \ $\nu_d\rightarrow{\bs v},\,{\bs v}={\bs v}+{\bs F}_f{\rm d}t,\, {\bs v}\rightarrow\nu_d$\\
\indent \ \ \ \ \tcv{enddo}\\
\indent \ \ \ \ update $\max(|{\dot{\bs v}}|)$\\
\indent \ \ \tcv{enddo}\\
\indent \ \ calculate $\rho_c$ and solve ${\bs F}_c$ in Fourier space\\
\indent \ \ \tcv{do} (each particle)\\
\indent \ \ \ \ $\nu_d\rightarrow{\bs v},\,{\bs v}={\bs v}+{\bs F}_c{\rm d}t,\, {\bs v}\rightarrow\nu_d$\\
\indent \ \ \tcv{enddo}\\
\indent \ \ update $\max(v_d),\max(|{\dot{\bs v}}|)$\\
\tcv{end}\\}
\caption{Pseudocode for subroutine {\tt \tcx{update\_vp}}.}
\label{fig.update_vp}
\end{figure}

The compensating coarse grid force ${\bs F}_c$ is globally computed by using a coarser- (usually by a factor of $R=4$) mesh by dimensional splitting -- the cubic distributed coarse density field $\rho_c$ is transposed (inter-image) and Fourier transformed (inner-image) in three consecutive dimensions. After the multiplication of memory-distributed force kernel $K_c$, the inverse transform takes place to get the cubic distributed coarse force field ${\bs F}_c$, upon which velocities are updated again.

For each type of velocity update, the collective operations are Equation(\ref{eq.v}), ${\bs v}={\bs v}+{\bs F}_{\rm total}$, and Equation(\ref{eq.nu}). We also update $\sigma^2_{\Delta}$ according to the new $v_d-v_c$. These steps are summarized in Figure \ref{fig.update_vp}.

After the updating of $\nu_d$ in $V_p$, we simply call {\tt \tcx{buffer\_v}} again to bring $\nu_d$ into the buffered state, such that the {\tt \tcx{update\_x}} in the next iteration will be done correctly.

For {\tt \tcx{checkpoint}},
if a desired redshift is reached, we execute the last drift step in the $({\rm DKKD})^n$ operation by {\tt \tcx{update\_xp}}, and call {\tt \tcx{checkpoint}} to save the disjoint state of $\{\rho_c,v_c,\chi_d,\nu_d\}$ on the disk. Other operations like run-time halo-finder or density projections are also done at this point.

Finally, in the {\tt \tcx{finalize}} subroutine
we destroy all the FFT plans and finish up any timing or statistics taken in the simulation.

\subsection{Memory layout}\label{ss.memory}
Here, we list the memory-consuming arrays and how they scale with different configurations of the simulation. We classify them into (1) arrays of particles, (2) coarse mesh arrays, and (3) fine mesh arrays.

\subsubsection{Arrays of particles}
These arrays comprise the majority of the memory usage, and contain checkpoint arrays $\chi_d$, $\nu_d$, $I_P$, and temporary arrays $\chi_d^*$, $\nu_d^*$, $I_P^*$. The former uses memory $\mathcal{M}=(3n_\chi+3n_\nu+n_I)P_{\rm max}\,{\rm byte}$, where
\begin{equation}\label{eq.npmax}
	P_{\rm max}=\left\langle P_{\rm local} \right\rangle \left( 1+\frac{2N_b}{N_t} \right)^3(1+\epsilon_{\rm image}),
\end{equation}
and $\left\langle P_{\rm local} \right\rangle$ is the average number of particles per image. The second term, proportional to $V_e/V_p$,  lets us store additional particles in $V_b$, and the third term $1+\epsilon_{\rm image}$ takes into account the inhomogeneity of $P_{\rm local}$ on different images. When each image models smaller physical scales, $\epsilon_{\rm image}$ should be set larger.

Temporary $\chi_d^*$, $\nu_d^*$, $I_P^*$ store particles only on tiles, and the particle number is set to be
\begin{equation}\label{eq.nptile}
	P_{\rm max}^*=\left\langle P_{\rm local} \right\rangle \left( \frac{1}{M_t} \right)^3(1+\epsilon_{\rm tile}),
\end{equation}
where $\epsilon_{\rm tile}$ controls the inhomogeneity on scales of tiles. Larger $M_t$ causes more inhomogeneity on smaller tiles, and $\epsilon_{\rm tile}$ can be much larger than $\epsilon_{\rm image}$; however, the term $M_t^{-3}$ decreases much faster. Practically, the majority memory is occupied by $\chi_d$, $\nu_d$, $I_P$.

Summarizing the above, we find that the memory usage per particle (bpp) ${\mathcal M}_P \equiv {\mathcal M} / \left\langle P_{\rm local} \right\rangle/{\rm byte}$, given $n_\chi=n_\nu=1$ and $n_I=0$, is
\begin{eqnarray}\label{eq.Mp_particle}
\nonumber\mathcal{M}_P^{\rm particle}=6\Big [ (1+2N_b N_t^{-1})^3(1+\epsilon_{\rm image} )\\
	+M_t^{-3}(1+\epsilon_{\rm tile})\Big ].
\end{eqnarray}
Because $N_t=N_c/(M_g M_t)$, we can minimize Equation (\ref{eq.Mp_particle}) by tuning $M_t$.

\subsubsection{Coarse mesh arrays}
On coarse mesh, $\rho_c$ (4-byte integers), $v_c$, and force kernel $K_c$ should always be kept. They are usually configured to be $R=4$ times coarser than fine grids and particle number density. They have use memories of $(1+3)\times 4\times(N_e M_t)^3+3\times 4\times(N_c/M_g)^3/2$ bytes per image, or
\begin{equation}\label{eq.Mp_coarse}
	\mathcal{M}_P^{\rm coarse}=P_c^{-1} \left [ 16\left( 1+\frac{2N_b}{N_t} \right)^3+6 \right ],
\end{equation}
where $P_c$ is the average number of particles per coarse cell. Other coarse-grid-based arrays are $\rho_c^*$, $F_c$, and pencil-FFT arrays. $\rho_c^*$ exists only on tiles; $F_c$ and pencil-FFT arrays can be equivalenced with other temporary arrays. Thus, the majority of the memory usage from coarse FFT arrays comes from Equation (\ref{eq.Mp_coarse}).

\subsubsection{Fine mesh arrays}
On the local fine mesh, only a force kernel array $K_f$ needs to be kept. It has memory of $3\times 4\times(RN_e M_t)^3/2$ per image, or
\begin{eqnarray}\label{eq.Mp_fine}
\nonumber	\mathcal{M}_P^{\rm fine}&=&6 P_f^{-1} N_t^{-3} \left ( 1+\frac{2N_b}{N_t} \right )^3 \\
					 &=& 6 R^3 P_c^{-1} N_t^{-3} \left ( 1+\frac{2N_b}{N_t} \right )^3,
\end{eqnarray}
where $P_f$ is the average number of particles per fine cell. For the fine mesh density arrays, force field arrays are temporary. Since $\chi_d^*$, $\nu_d^*$, coarse force arrays, and pencil-FFT arrays are also temporary and are not used in any calculation simultaneously, they can be overlapped in memory by using equivalent statements.

\begin{table}[]
\centering
\caption{Memory layout for a certain configuration}
\label{t.memory}
\begin{tabular}{llrrr}
\hline
& & \multicolumn{3}{c}{Memory usage}\\
\cline{3-5}
Type     & Array & /GB & /bpp & Percentage \\
\hline
Particles  & $\chi_d$, $\nu_d$   & 29.9   & 8.24   & 83.8\%     \\
           & $\chi_d^*$, $\nu_d^*$  & \underline{3.16}   & \underline{0.872}  & \underline{8.87\%}     \\
           & $I_P$, $I_P^*$        & 0      & 0      &    0\%     \\
           & Subtotal               & 33.0   & 9.12   & 92.7\%     \\
\hline
Coarse mesh& $\rho_c$               & 0.296  & 0.0818 & 0.831\%    \\
           & $v_c$                  & 0.889  & 0.245  & 2.49\%     \\
           & $K_c$                  & 0.340  & 0.0939 & 0.954\%    \\
           & $F_c$                  & (0.690)  & (0.190)  & (1.94\%)    \\
           & $\rho_c^*$             & 0.0140 & 0.00388  & 0.0394\% \\
           & Pencil-FFT             & (0.454)  & (0.125) & (1.27\%)      \\
           & Subtotal               & 1.55   & 0.429 & 4.36\%     \\
\hline
Fine mesh  & $K_f$                  & 1.06   & 0.292 & 2.97\%      \\
           & $F_f$                 & (1.63) & (0.450) & (4.57\%)     \\
           & Fine-FFT               &   (1.41) & (0.389) & (3.96\%)     \\
           & Subtotal               & 1.06   & 0.292 & 2.97	\%      \\
\hline
Total &                      & 35.6 & 9.84 & 100\%\\
\hline
\multicolumn{2}{l} {Optimal limit} & 21.7 & 6 & 61.0\%\\
\hline
\end{tabular}
\end{table}

\begin{figure*}[]
\centering
  \includegraphics[width=0.75\linewidth]{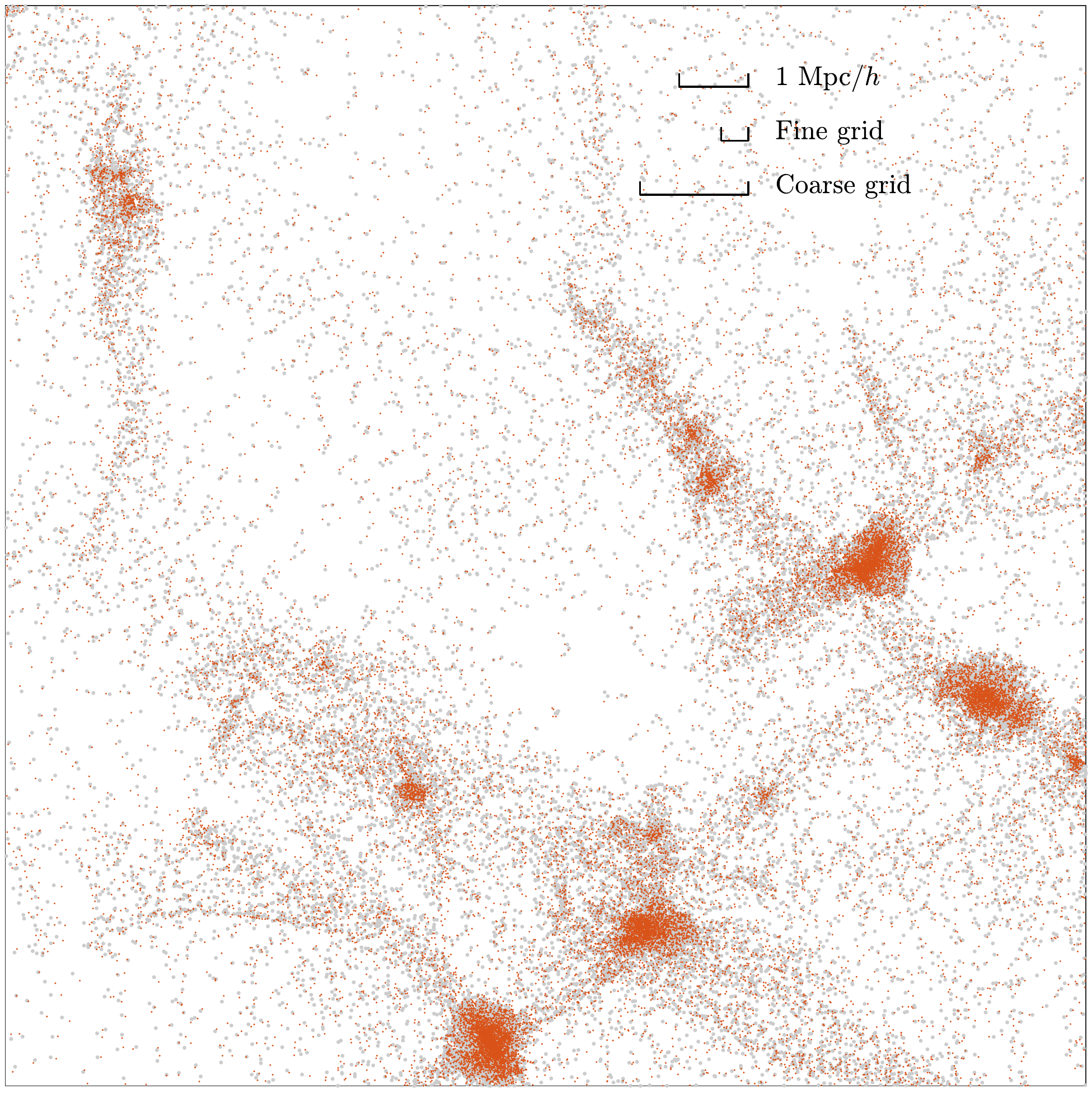}
 \caption{Offset in the particle distribution induced by an integer-1 based algorithm (x1v1). In S512's subregion of volume $15.625^2\times 3.90625\ ({\rm Mpc}/h)^3$, particles from {\tt CUBEP3M} (the larger gray dots in the background) and {\tt CUBE}-x1v1 (the smaller red dots) are projected onto the plane of $(15.625\ {\rm Mpc}/h)^2$. The comparing rules show 1 ${\rm Mpc}/h$, for fine and coarse grids respectively, and position resolution of x1v1 is 1/64 of a fine grid.}
\label{fig.particles}
\end{figure*}

\subsubsection{Compare with traditional algorithms}
To illustrate the improvement of memory usage, we refer to the TianNu simulation \citep{2017NatAs...1E.143Y} run on the Tianhe-2 supercomputer, which used a traditional $N$-body code {\tt CUBEP3M}. TianNu's particle number (shown in Table \ref{t.sim}) is limited by memory per computing node -- for each computing node, an average of $576^3$ neutrino particles and $288^3$ CDM particles are used, and consumes $\mathcal{M}=40\,{\rm GB}$\footnote{Additional memory is used for OpenMP parallelization and for particle IDs to differentiate different particle spices.}, or about $\mathcal{M}_P=186$. A memory-efficient configuration of {\tt CUBEP3M} by using large physical scales and at costs of speed, still uses about $\mathcal{M}_P=40$.

If the same amount of memory is allocated to {\tt CUBE}, we can set parameters as $n_\chi=n_\nu=1$, $N_c/M_g=384$, $N_b=6$, $\epsilon_{\rm image}=5\%$, $\epsilon_{\rm tile}=200\%$ and thus $\left\langle P_{\rm local}\right\rangle=1536^3$. Setting $M_t=3$ (27 tiles per image) minimizes $\mathcal{M}$ and uses about $\mathcal{M}=35.6\,{\rm GB}$, corresponding to $\mathcal{M}_P=9.84$. This can be done on most of the supercomputers, even modern laptops.

Table \ref{t.memory} shows the memory consumption for this test simulation. The memory-consuming arrays are listed and classified into the three types above mentioned, and their memory usages are in units of GB ($10^9$ byte), bpp, and their percentage of the total memory usage. The parenthesized numbers show overlapped memory, which is saved by equivalencing them with the underscored numbers. There are other unlisted variables that are memory-light, and can also be equivalenced with the listed variables. 

In the bottom of Table \ref{t.memory} we stress that the optimal memory usage is 6 bpp, or 21.7 GB, 61\% of the actual $\mathcal{M}$. The dominating departure from this limit is that $\chi_d$ and $\nu_d$ already occupy 8.24 bpp, which come from the $(1+2N_b/N_t)^3$ term of Equation (\ref{eq.Mp_particle}). All other variables occupy an additional 8\%. On modern supercomputers, the memory per computing node is usually much larger, and by scaling up the number of particles per node, the buffer ratio $N_b/N_t$ will be lowered and we can approach closer to the 6 bpp limit.

\begin{table}[]
\centering
\caption{Simulation configurations}
\label{t.sim}
\begin{tabular}{lrrrrr}
\hline
& \multicolumn{5}{c}{Configurations}\\
\cline{2-6}
Name  & $N_{\rm node}$ & $L/({\rm Mpc\,}h^{-1})$ & $z_i$ & $N_p$ & $m_p/\Msun$ \\
\hline
S512   & 8     & 200   & 49   & $512^3$  & $7.5\times 10^9$    \\
S256   & 1     & 80    & 49   & $256^3$  & $3.8\times 10^9$    \\
S2048S & 64    & 400   & 49   & $2048^3$ & $9.4\times 10^8$    \\
S2048L & 64    & 1200  & 49   & $2048^3$ & $2.5\times 10^{10}$ \\
\hline
TianNu & 13824 & 1200  & 100  & $6912^3$ & $6.9\times 10^8$\\
       &       &       & 5    & $13824^3$& $3.2\times 10^5$\\
TianZero & 13824 & 1200 & 100 & $6912^3$ & $7.0\times 10^8$\\
\hline
\end{tabular}
\end{table}

\begin{figure}[]
\centering
  \includegraphics[width=0.99\linewidth]{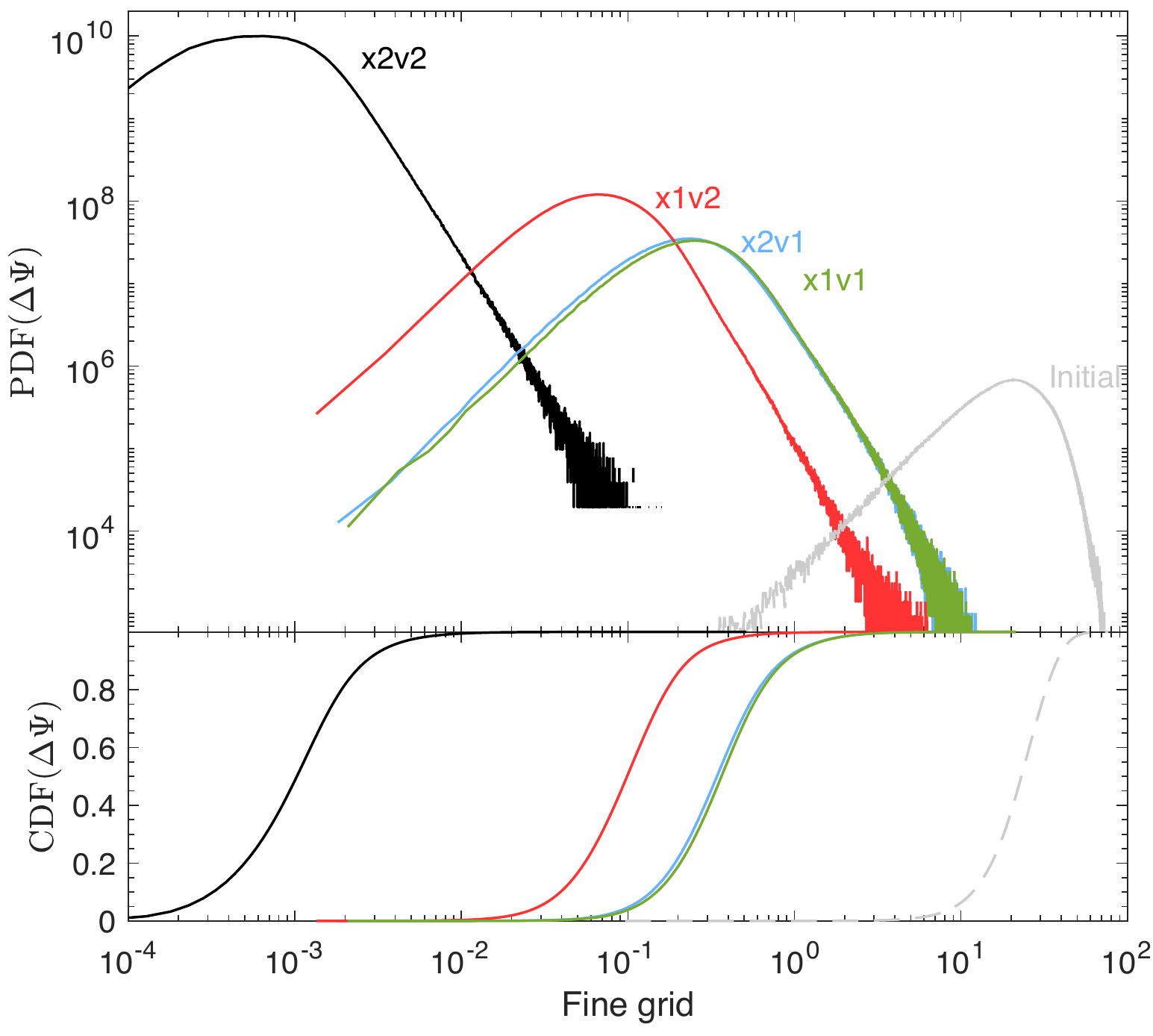}
 \caption{Statistics of the error in particle displacement $\Delta{\bs\Psi}$ induced by integer-based algorithms in {\tt CUBE}. The PDF and CDF of $|\Delta{\bs\Psi}|$ from S256 are shown in the upper and lower axes, as functions of fine grids. Black, red, blue, and green correspond to x2v2, x1v2, x2v1, and x1v1, respectively. The gray lines, marked with ``initial'', show the distribution of the actual displacement of particles in {\tt CUBEP3M}, $|{\bs\Psi}_0|$, which is orders of magnitudes larger than $|\Delta{\bs\Psi}|$. }
\label{fig.dsp}
\end{figure}

\section{Accuracy}\label{s.results}
We run a group of simulations to test the accuracy of {\tt CUBE}. We use the same seeds to generate the same Gaussian random fields in the initial condition generators of {\tt CUBEP3M} and {\tt CUBE}, and then they produce initial conditions of their own formats. Then, the main $N$-body codes run their own initial conditions to redshift $z=0$. We use the same force kernels as {\tt CUBEP3M} without PP force. Note that near the find grid scales, it is possible to enhance the force kernel to better match the nonlinear power spectrum predictions; however, we use the conservative mode of {\tt CUBEP3M} in this paper. An extended PP force and an unbiased force matching algorithm will be added to {\tt CUBE}. The power spectrum studies are presented in \citet{2013MNRAS.436..540H}, while here we focus on the cross-correlations between different integer-based methods.

First, by using different configurations -- different numbers of computing nodes, box sizes, particle resolutions, different number of tiles per node/image, etc., we find that by using 2-byte integers for both positions and velocities ($n_\chi=n_\nu=2$, or x2v2) allows {\tt CUBE} to give exact results compared to {\tt CUBEP3M}. So if sufficient memory is provided, one can always use x2v2 to get exact results as {\tt CUBEP3M}, and the optimal memory limit of this case is 12 bpp, which is still much lower than traditional methods. Next, we focus on the accuracy of the other three cases -- x1v2, x2v1, and x1v1.

We list the names and configurations of the simulations used in Table \ref{t.sim}, where $N_{\rm node}$, $L$, $z_i$, $N_p$, and $m_p$ are respectively the number of computing nodes used, the length of the side of the box, initial redshift, total number of particles, and particle mass. These configurations are run by {\tt CUBEP3M}, x2v2, x1v2, x2v1, and x1v1 versions of {\tt CUBE} with the same initial seeds. Using different numbers of tiles per image gives the exact same results. We also list the configurations for TianNu and TianZero \citep{2017NatAs...1E.143Y,2017RAA....17...85E} simulations as a reference.

\begin{figure*}
\centering
  \includegraphics[width=1.0\linewidth]{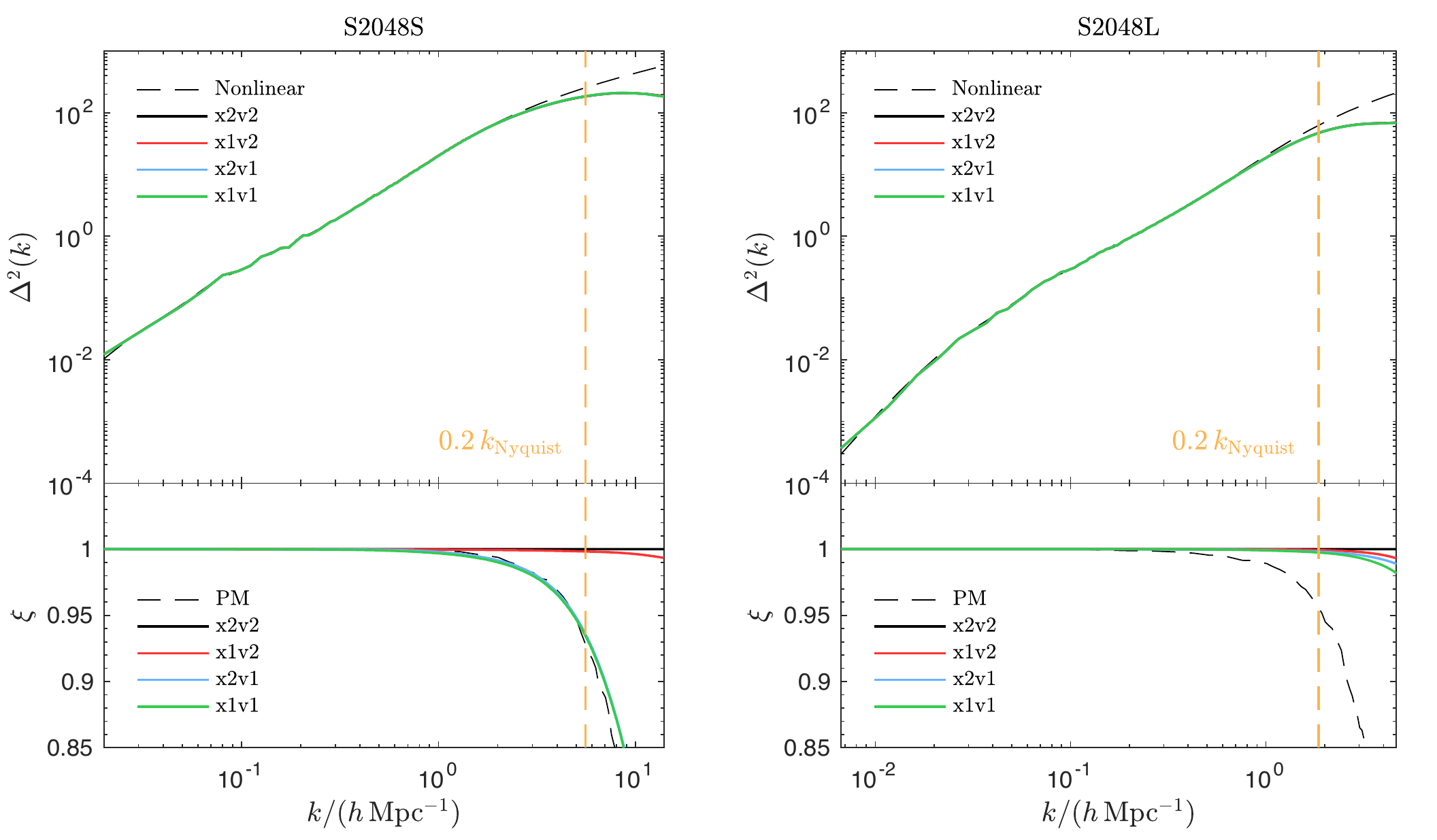}
 \caption{Dimensionless power spectra $\Delta^2(k)$ and the cross-correlations $\xi(k)$ (lower axes) with respect to x2v2 in simulations S2048S and S2048L. The four solid lines show the results from \{x2v2,x1v2,x2v1,x1v1\}. In the lower panels, the dashed curves show the decorrelations given by the PM algorithm, measured by cross-correlating different resolutions of {\tt CUBE}-x2v2. The vertical orange dashed lines show the scale $k=0.2\, k_{\rm Nyquist}$.}
\label{fig.ccc}
\end{figure*}
\subsection{Power spectrum}

\subsection{Displacement of particles}
In S512, we zoom-in on a small region of $15.625^2\times 3.90625\ ({\rm Mpc}/h)^3$ and compare the particle distribution between {\tt CUBEP3M} and {\tt CUBE}-x1v1 in Figure \ref{fig.particles}. For clarity, {\tt CUBEP3M} particles are marked with the larger gray dots, whereas the smaller red dots are {\tt CUBE}-x1v1 particles overplotted onto them. The 1 ${\rm Mpc}/h$, fine grid, and coarse grid scales are shown in the figure. The position resolution of particles in {\tt CUBE}-x1v1 is 1/256 of a coarse grid, or 1/64 of a fine grid.

To quantify the offset in the final particle distributions, we use PIDs to track the displacement ${\bs \Psi}({\bs q})\equiv{\bs x}-{\bs q}$ of every particle \citep{2017PhRvD..95d3501Y}, where ${\bs x}$ and ${\bs q}$ are Eulerian and Lagrangian coordinates of the particle. Then, we calculate the absolute value of the offset vector
\begin{equation}\label{eq.offset}
	\Delta\Psi\equiv|{\bs\Psi}_i-{\bs\Psi}_0|.
\end{equation}
Here, ${\bs\Psi}_0$ stands for {\tt CUBEP3M} and subscript $_i$ can stand for x2v2, x1v2, x2v1, or x1v1. The PDFs and CDFs of $\Delta\bs\Psi$ in S256 are shown in Figure \ref{fig.dsp}. Results from x2v2, x1v2, x2v1, or x1v1 are in black, red, blue, and green respectively. The results from absolute displacement of particles (by replacing ${\bs\Psi}_i$ with ${\bs q}$ in Equation (\ref{eq.offset})) are shown in gray for comparison.

For x2v2, almost all particles are accurate up to 1/100 of a fine grid, and the worst particle is $\sim 1/10$ of a fine grid away from its counterpart in {\tt CUBEP3M}. The difference is caused by round-off errors and is negligible in physical and cosmological applications. The accuracy of x1v2 is between x2v2 and x1v1, and x2v1 gives only a minor improvement from x1v1. We also run a simulation with the same number of particles but with $L=600\ {\rm Mpc}/h$ ($m_p=1.2\times 10^{12}\Msun$), and find that the accuracy of x1v2 is in turns between x2v1 and x1v1. We interpret that, in this latter case, particles have lower mass resolution, so they move slower and need higher position resolution but need lower velocity resolution, thus x2v1 outperforms x1v2.

S2048S and S2048L are two simulations with $2048^3$ particles in small ($L=400\ {\rm Mpc}/h$) and large ($L=1200\ {\rm Mpc}/h$) box sizes. We compare their accuracy by their power spectra and their cross-correlations with {\tt CUBEP3M} at $z=0$. The physical scale of S2048S is designed such that the particle mass resolution $m_p$ is comparable to TianNu and TianZero simulations (their parameters are also listed in Table \ref{t.sim}). On the other hand, S2048L focuses on larger structures, on which scale one can study weak gravitational lensing, BAO \citep{2005ApJ...633..560E}, and its reconstruction \citep{2007ApJ...664..675E,2017ApJ...841L..29W}, etc.

For each simulation the particles are firstly cloud-in-cell (CIC) interpolated onto the fine mesh grid, and from the density field $\rho$ we define the density contrast $\delta\equiv \rho/\left\langle \rho \right\rangle-1$. We define the cross-power spectrum $P_{\alpha\beta}(k)$ between two fields $\delta_\alpha$ and $\delta_\beta$ ($\delta_\alpha=\delta_\beta$ for auto-power spectrum) in Fourier space as
\begin{equation}\label{eq.cps}
	\left\langle \delta_\alpha^\dagger({\bs k})\delta_\beta({\bs k}') \right\rangle=
    (2\pi)^3 P_{\alpha\beta}(k){\bs \delta}_{\rm 3D}({\bs k}-{\bs k}'),
\end{equation}
where ${\bs \delta}_{\rm 3D}$ is the 3D Dirac delta function. In cosmology we usually consider the dimensionless power spectrum $\Delta^2_{\alpha\beta}(k)\equiv k^3 P_{\alpha\beta}(k)/(2\pi^2)$. The cross-correlation coefficient is defined as
\begin{equation}\label{eq.ccc}
	\xi(k)\equiv P_{\alpha\beta}/\sqrt{P_{\alpha\alpha}P_{\beta\beta}}.
\end{equation}

In the upper two panels of figure \ref{fig.ccc} we show the power spectra of {\tt CUBE}. In both plots of S2048S and S2048L the four solid curves of different colors show the results of x2v2, x1v2, x2v1, and x1v1, and they almost overlapped with each other. The dashed curves are the nonlinear prediction of the matter power spectrum by CLASS \citep{2011JCAP...07..034B}.

We label $k=0.2\,k_{\rm Nyquist}$ as vertical dashed lines, where $k_{\rm Nyquist}$ is the scale of fine mesh grids, and the scale of average particle separations. On this scale, the power spectra are offset from nonlinear predictions by at least 20\%. This error is from the PM algorithm and one has to increase the resolution of the simulation to correct these offsets. We do not plot {\tt CUBEP3M} because we found that {\tt CUBEP3M} and x2v2 of {\tt CUBE} produce same results. Thus, the differences between {\tt CUBEP3M} and different integer formats of {\tt CUBE} are negligible compared to the error of the PM algorithm.

In the lower parts of these four panels we study the cross-correlations. We compare everything with {\tt CUBE}-x2v2, to emphasize the decorrelation (i.e. $1-\xi(k)$) by different integer formats. Note that x2v2 is perfectly correlated with {\tt CUBEP3M}. In the lower two panels of figure \ref{fig.ccc} the solid curves show the decorrelation $\xi$ of x1v2, x2v1, and x1v1. The higher resolution (S2048S) in general comes with more decorrelations, and x1v2 cross-correlates with x2v2 better than the other two cases. In order to quantify the PM error in terms of decorrelations, we run x2v2 simulations with the same initial conditions, but 8 times more particles and cells, and measure the decorrelation caused by coarser resolutions. These results are shown as dashed curves (labeled ``PM''). We conclude that the cross-correlation of x1v1, in all cases, is not worse than the PM errors.

To summarize from Figure \ref{fig.ccc}, in terms of either power spectrum deviation or cross-correlation, the error induced by information optimization (even for x1v1) is lower than the error from the PM algorithm, and we can safely use x1v1 for most of the LSS studies.

\section{Discussion and conclusion}\label{s.discussion}
We present a parallel, information-optimized $N$-body algorithm. This open-source code, {\tt CUBE}, has recently been used in many studies of LSS, e.g. \cite{2017PhRvD..95d3501Y,2017ApJ...841L..29W,2017MNRAS.469.1968P}. It requires very low memory usage, approaching 6 bpp.

The accuracy of this code is adjustable in that we can choose 1-byte/2-byte integers separately for positions and velocities of particles. In the case of using 2 byte integers for both positions and velocities (``x2v2'', and memory can be 12 bpp), the algorithm gives the exact results given by traditional $N$-body algorithms. Note that the results are exactly the same in that they not only produce the same physical statistics of LSS, but also the same error (not physical) from the PM algorithm near Nyquist frequencies. In other words, the positions and velocities of each particle are exact. In practice, we only require that the errors from information optimization is much lower than the errors from the PM algorithm. In Figure \ref{fig.ccc} we see that this is the case even for the most memory-efficient configuration, x1v1. This shows that in most LSS studies, when our scales of interest are smaller than $k\simeq 0.2\,k_{\rm Nyquist}$, six 1-byte fixed point numbers contain sufficient information of every $N$-body particle.

Another benefit of this algorithm is LSS simulations with neutrinos, although the neutrino modules of {\tt CUBE} are in development. Neutrinos have a high velocity dispersion and move much faster than CDM, and their small-scale errors are dominated by their Poisson noise. We expect that, compared to CDM, x1v1 gives less power spectrum deviation to neutrinos, as they behave more Gaussian and are less clustered. LSS-neutrino simulations, like TianNu, can contain much more (for TianNu, 8 times more) neutrino $N$-body particles than CDM, which dominate the memory. For a TianNu-like simulation, one can safely use x1v2 or x2v2 for CDM and x1v1 for neutrinos, and the memory usage can approach 6 bpp. This allows much more particles to be included in the simulation and can lower the Poisson noise of neutrinos prominently. A 1-byte PID per particle increases minor memory usage and can differentiate eight kinds of particles, or we can store different particles in different arrays without using PID.

We did not include PP force in {\tt CUBE} and {\tt CUBEP3M} in this paper. If one wants to focus on smaller scales, like halo masses and profiles, an extended PP forces, which act up to adjacent fine cells, should be taken into account. The memory consumption for PP force is only local and is negligible compared to particles. In these cases where even x2v2 is used, a 12 bpp memory usage is still much lower than that of traditional $N$-body algorithms.

Traditional $N$-body codes consume considerable memory while performing relatively light computations. {\tt CUBE} is designed to optimize the efficiency of information and the memory usage in $N$-body simulations. {\tt CUBE} is written in Coarray Fortran -- concise Coarray features are used instead of complicated MPI -- and the code itself is much more concise than {\tt CUBEP3M} for future maintenance and development. The next steps are optimization of the code and adapting it for various kinds of heterogeneous computing systems, e.g. MIC and GPUs. Optimizing the velocity storage may further improve the accuracy of x1v1, and whose effects on neutrino-LSS simulation are yet to be discovered.

\acknowledgements
We acknowledge funding from NSERC.
H.R.Y. thanks Derek Inman for many helpful discussions. We thank the anonymous referee for many helpful suggestions that improve the paper.
The simulations were performed on the GPC supercomputer at the SciNet HPC Consortium.
The code is publicly available on github.com under {\tt yuhaoran/CUBE}.

\bibliographystyle{aasjournal}
\bibliography{haoran_ref}

\end{document}